\documentclass[seceq]{ptptex}

\usepackage{epsfig}

\title{Voids in the distribution of galaxies and the Cosmological constant}

\author{
Roland \textsc{Triay}$^{1}$ and Henri-Hugues \textsc{Fliche}$^{2}$
}

\inst{
$^1$ Centre de Physique Th\'eorique\footnote{Unit\'e Mixte de Recherche (UMR 6207) du CNRS, et des universit\'es Aix-Marseille I, Aix-Marseille II et du Sud Toulon-Var. Laboratoire affili\'e \`a la FRUMAM (FR 2291).} CNRS Luminy Case 907\\ 13288 Marseille Cedex 9, France\\
$^2$ LMMT\footnote{UPRES~EA 2596}, Fac. des Sciences et Techniques de St J\'er\^ome\\
av. Normandie-Niemen, 13397 Marseille Cedex 20, France

}

\abst{
With the motivation in mind to evaluate the contribution of the cosmological constant $\Lambda$ on the foam like patterns formation process in the distribution of galaxies, we investigate the Newtonian dynamics of a spherical void embedded in an uniform medium which undergoes a Hubble expansion.  We use a covariant approach for deriving the evolution with time of the shell (S) acting as a boundaries condition for the inside and outside media. As a result,  with the usual values for the cosmological parameters, S expands with a huge initial burst that freezes up to matching Hubble flow.  With respect to Friedmann comoving frame, its magnification increases nonlinearly with $\Lambda$, with a maximal growth rate at redshift $z\sim 1.7$. The velocity field inside S shows an interesting feature which enables us to disentangle a spatially closed from open universe. Namely, the void region are swept out in the first case, what can be interpreted as a stability criterion.}

\begin{document}

\maketitle

\section{Introduction}
The understanding of the foam like patterns with large empty regions in the distribution of galaxies\cite{SoneiraPeebles78, JoeveerEinasto78,Einasto02,WeygaertEtal04} has become an important challenge for the formation theory of cosmological structures\cite{Peebles01}. Our motivation for understanding the dynamics of a single empty void in an expanding universe is twofold. According to Schwarzschild solution, the cosmological constant $\Lambda$ that is required for the interpretation of high redshift data makes repulsive the gravity at a given critical distance\cite{Triay05}, what forces us to understand the underlying effect on the dynamics of voids since it does not limit solely to a pure chronological effect as in a Friedmann model. Additionally, the usual approaches based on gravitational collapse of density excess in which perturbations theory can be successfully applied from Zeldovich's approximation \cite{Zeldovich68} are not valid for voids because they never evolve within a linear regime\cite{BenhamidoucheEtal99}, what motivates us to look for exact solutions. We limit our investigation to a newtonian approach \cite{TriayFliche06a} as a first step of a more general N-bodies treatment, although the related results must be interpreted with care when cosmological distances are involved\cite{TriayFliche07}. Relativistic treatments of an analogous problem (an empty bubble as singularity) have been performed \cite{Sato82, MaedaEtal82, MaedaSato83, SatoMaeda83, Maeda86} but with a different aim.

\section{A void model in an expanding universe}
For the investigation of structures formation in an isotropic expanding medium, instead of position $\vec{r}=a\vec{x}$, it is convenient to use {\it reference coordinates\/} $(t, \vec{x})$ related to the expansion factor  $a=a(t)$; let us choose a normalization condition at a given date $a_{\circ}=a(t_{\circ})=1$. Hence, the density $\rho=\rho(\vec{r},t)$ and velocity fields $\vec{v}=\vec{v}(\vec{r},t)$ of the cosmic fluid translate in the appropriated {\em reference frame\/} by
\begin{eqnarray}\label{VetG}
\rho_{\rm c}=\rho a^{3},\qquad
\vec{v}_{\rm c}=\frac{{\rm d}\vec{x}}{{\rm d}t}=\frac{1}{a}\left(\vec{v}-H\vec{r}\right),\qquad H=\frac{\dot{a}}{a}
\end{eqnarray}
According to the standard world model,  we assume that the gravitational sources behave as dust. The  motions of such a pressureless medium are constraint by Euler-Poisson-Newton (EPN) equations system. For accounting of the cosmological expansion, the function $a(t)$ must verify a Friedmann type equation
\begin{equation}\label{chrono}
H^{2}=\frac{\Lambda}{3}-\frac{K_{\circ}}{a^{2}}+\frac{8\pi{\rm G}}{3} \frac{\rho_{\circ}}{a^{3}}\geq 0,\qquad
K_{\circ}=\frac{8\pi{\rm G}}{3} \rho_{\circ}+\frac{\Lambda}{3}-H_{\circ}^{2}
\end{equation}
what depends on two parameters chosen among $\Lambda$, $\rho_{\circ}$ and $K_{\circ}$ (which stands for an integration constant in Friedmann differential equation). It is important to mention that at this step, excepted $G$ and $\Lambda$, these variables stand merely for a coordinates choice. The formal expression $t\mapsto a$ is derived as reciprocal mapping of a quadrature from eq.\,(\ref{chrono}), what gives the reference frame chronology. It turns out that EPN equations system shows two trivial solutions~:
\begin{enumerate}
\item Newton-Friedmann (NF) solution, defined by $\rho_{\rm c}=\rho_{\circ}$ and $\vec{v}_{\rm c}=\vec{0}$, which accounts for a uniform distribution of dust. According to eq.\,(\ref{VetG},\ref{chrono}), $a$ and $H$ recover now their usual interpretations in cosmology as  {\em expansion parameter\/} and {\em Hubble parameter\/} respectively, $\rho_{\circ}=\rho a^{3}$ identifies to the density of sources in the comoving space and $K_{\circ}$ interprets in GR as its scalar curvature\footnote{We limit our investigation to motions which do not correspond to cosmological bouncing solutions, what requires the constraint
$
K_{\circ}^{3}<\left(4\pi{\rm G}\rho_{\circ}\right)^{2}\Lambda
$
to be fulfilled, according to analysis on roots of third degree polynomials.}, $k_{\circ}=K_{\circ}H_{\circ}^{-2}$ being its dimensionless measure \footnote{A dimensional analysis of Eq.\,(\ref{chrono}) shows that the Newtonian interpretation of $k_{\circ}$ corresponds to a dimensionless binding energy for the universe. The lower $k_{\circ}$ the faster the cosmological expansion.}. Such a model depends on a scale parameter $H_{\circ}=H(t_{\circ})$ and two dimensionless parameters $\Omega_{\circ}=\frac{8}{3}\pi G H_{\circ}^{-2}\rho_{c}$ and $\lambda_{\circ}=\frac{1}{3}\Lambda H_{\circ}^{-2}$. 
\item  Vacuum (V) solution, defined by $\rho_{\rm c}=0$ and $\vec{v}_{\rm c}=\left( \frac{\Lambda}{3}-H \right) \vec{x}$.
\end{enumerate}

\begin{figure}[htbp]
\begin{center}
\epsfig{file=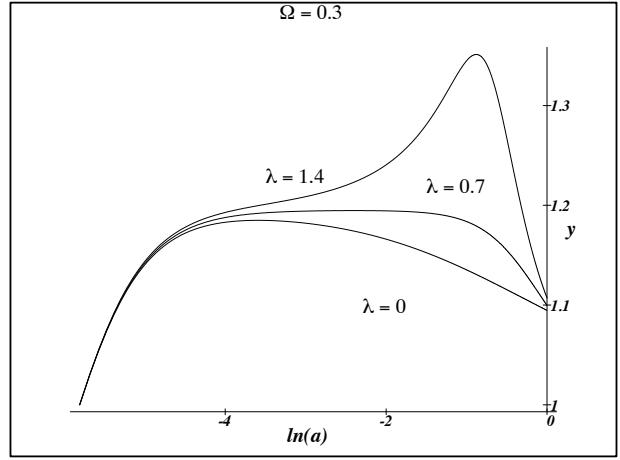,width=3.5in}
\caption{The corrective factor to Hubble expansion.}
\label{Fig0}
\end{center}
\end{figure}

The dynamics of a spherical void surrounded by an uniform distribution of gravitational sources is obtained by using a covariant formulation of Euler-Poisson equations system \cite{Souriau70,DuvalKunzle78}, which allows us to stick together the local V and NF solutions. We assume that their common border is  a {\it material shell\/} (S) with a negligible tension-stress characterized  by a (symmetric contravariant) mass-momentum tensor
\begin{equation}\label{Ts}
T_{\rm S}^{00}=(\rho_{\rm S})_{c},\quad
T_{\rm S}^{0j}=(\rho_{\rm S})_{c}v_{c}^{j},\quad
T_{\rm S}^{jk}=(\rho_{\rm S})_{c}v_{c}^{j}v_{c}^{k}
\end{equation}
The NF-kinematics, which applies to the outer part of S for describing the cosmological expansion (Hubble flow), shows two distinct behaviors characterized by the sign of $k_{\circ}$~: $H$ decreases with $a$ by reaching its limiting value $H_{\infty}$ either upward ($k_{\circ}\leq 0$) or downward ($k_{\circ}>0$) from a minimum
 \begin{equation}\label{extremum}
H_{m}=H_{\infty}\sqrt{1-\frac{K_{\circ}^{3}}{\Lambda\left(4\pi{\rm G} \rho_{\circ}\right)^{2}}}<H_{\infty}=\lim_{a\to\infty}H=\sqrt{\frac{\Lambda}{3}}
\end{equation}
at (epoch) $a=4\pi{\rm G}\rho_{\circ}K_{\circ}^{-1}$, what defines a  {\it loitering period\/}.
The expansion velocity of the shell S with respect to its centre reads $\vec{v}=y H \vec{r}$, where the corrective factor $y$ to Hubble expansion versus $a$ is given in Fig.\,\ref{Fig0}. It results from a vanishing initial expansion velocity at  $a=0.003$, and has been evaluated for three world models defined by $\Omega_{\circ}=0.3.$ and  $\lambda_{\circ}=0$,  $\lambda_{\circ}= 0.7$, $\lambda_{\circ}=1.4$.
It can be noted that S expands faster than Hubble expansion ($y>1$) with no significant dependence on $\Lambda$ at early stages of its evolution. However, the presence of $\Lambda$ magnifies its expansion (and hence its size) later during its evolution, and the larger the $\lambda_{\circ}$ ({\it i.e.\/}, $k_{\circ}$) the major this effect. Such an effect shows its maximum at redshift $z\sim 1.7$ and then decreases with time for reaching asymptotically Hubble flow\footnote{This bump on the diagram results from the cosmological expansion which undergoes a loitering period while the void expands, see Eq.\,(\ref{extremum}). The present epoch $t=t_{\circ}$ appears quite peculiar because of the relative proximity of curves, but it is solely an artifact due to the choice of the scale.}. Todays, its magnitude, which is of order of 10 percent of Hubble expansion, contributes possibly to the peculiar motion of our galactic region away from the Local Void\cite{TullyEtal08,Tully08a,Tully08b}. 

\section{Discussion}
We have proved that a spherical void stands for a \underline{global solution} of Euler-PoissonÕs equations, which provides us with an hint on $\Lambda$ effect on the dynamics of voids in the distribution of galaxies. More realistic shapes cannot be treated analytically for taking into account the active role of such a structure in the description of gravitational field, unless by numerical methods. The reason is that the radial symmetry is required  for the boundary region (S) in order to preserve the isotropy of  Friedmann model (also required in GR), otherwise another background model must envisaged.

An interesting property of the velocity field inside S enables us to disentangle a spatially closed universe ($k_{\circ}>0$) from an open one if  inside S few galaxies are present so that to not contribute significantly to the gravitational fields. Indeed, in the reference frame, a test-particle moves toward the centre of S but if $k_{\circ}>0$ then it starts moving toward its border at (date) $a=\frac{8}{3} \pi{\rm G}\rho_{\circ}K_{\circ}^{-1}$. Such a property of sweeping out the void region (due to $H<H_{\infty}$) interprets as a stability criterion for void regions.

\section*{Acknowledgements}
I would like to thank Prof. Masakatsu Kenmoku for his hospitality and successful organization of the conference.

%

\end{document}